# Termination shock as a source of unusual solar radio bursts

Valery Fomichev and Gennady Chernov; gchernov@izmiran.ru

Puskov Institute of Terrestrial Magnetism, Ionosphere and Radio Wave Propagation of Russian Academy of Science (IZMIRAN),

Kaluzhskoe Hwy 4, Troitsk, Moscow, 108840, Russia;

## Abstract

Using centimeter-wave and decimeter-wave solar radio spectral observations of the flares of November 18, 2003 and September 12, 2004, we have discussed two type II–like bursts at the meter waves. The radio bursts show that the ordinary frequency drift from high to low frequencies slows down and stops, and a frequency drift from low to high frequencies appears. An analysis of all data on the corresponding flares provides evidence of formation of quasi-standing fast-mode shocks (termination shocks, TS). TS are able to generate energetic electrons, responsible for the appearance of new sources of hard X-ray radiation and generation of fast radio bursts (spikes), fibers and zebra-structures. The sources of the radio emission bands with the unusual frequency drift are situated above the top of the post-flare loops (lower TS) or are connected with the erupting prominence or coronal mass ejection (CME, upper TS). Estimations of the critical Mach numbers for the ordinary plasma parameters in the solar flares give the values $M_{cr} = 1.1-1.3$ easily realized in the flare events. The conditions necessary for generation of unusual radio bursts are likely to occur in the helmet-shaped magnetic structures in the solar corona.

*Key words*: Sun: radio emission — radio spectrograph — methods  — acceleration of particles — shock waves — solar flares

## 1. Introduction

The present-day classification of the solar radio bursts is based on observations of their dynamic spectra, generally in the decimeter and meter wavelength ranges (Zheleznyakov 1964; Kundu 1965). However, the dynamic spectra of the solar radio bursts in a broader frequency range reveal unusual radio bursts that are difficult to attribute to any of the five known classical types and to their fine structures (Chernov 2011). Examples of radio bursts of this type are the drifting bands in the microwave range, which are rather like the type II radio bursts in the meter wavelength range. The type II solar radio bursts are known to be initiated by shock waves arising because of solar flares and mass ejections. In the classical type II bursts, the emission bands drift from high to low frequencies. In the frames of the common plasma generation mechanism this corresponds to the propagation of shock waves outwards in the solar corona from high to low electron density. The typical velocity of the shock ranges from 500 to 2500 km/s.

Another example is the dynamic spectrum of the unusual and rare U-shaped type II radio burst associated with the March 28, 1980 flare (Markeev et al. 1983). In this event, the type II burst, first, displayed an ordinary negative drift (i.e drift from high to low frequencies) and then, after having reached the plateau level, a positive drift. In that paper, it was suggested that the shock wave generated by the flare propagated through ducting and repeated reflections, which take place within a large-scale coronal loop or an arcade of loops, thus resulting in reversed frequency drift in the type II emission. Some observations made from Culgoora heliograph and Clark lake array also showed that in few events the shock may undergo refraction and propagate along a curved path, guided by magnetic fields (Kai 1969; Smerd 1970; Dulk et al., 1971; Gergely & Kundu 1981)

Type II radio bursts without frequency drift were also discussed in (Mann et al., 2006, 2009; Warmuth et al., 2009), where the radio emission was hypothetically associated with the appearance of the termination shock waves (TS). Appearance of such TS is predicted in the reconnection model of flare energy release (Kosugi & Somov 1998; Priest & Forbes 2000) where hot reconnection outflow jets arise surrounded by a system of standing slow-mode shocks in the corona. In the jets, standing fast-mode shocks (called termination shocks [TSs]) are expected somewhere between the diffusion region (DR) and the erupting prominence (upper TS), and between the DR and the top of the postflare loops (lower TS) if the speed of these outflows is super- Alfvenic. Fast-mode shocks are usually able to generate energetic electrons and thus possibly also hard X-ray and radio emission. The possibility of acceleration in TS was first suggested by Tsuneta & Naito (1998). But for a long time, the existence of TS in the solar atmosphere had been a hypothesis until the observation evidence (e.g. Aurass & Mann, 2004; Aurass et al., 2006) and numerical simulations of the physical processes in solar flares (Forbes, 1986; Shibata et al., 1995) appeared. In a number of papers the electron drift acceleration at the TS was analyzed in details in the reconnection geometry (Mann et al., 2006, 2009; Warmuth et al., 2009; Aurass et al. 2013), and it was shown a possibility to explain the data on X-ray and radio emission in some solar flares.

In this connection, it should be noted also the two-dimensional resistive magnetohydrodynamic (MHD) simulation of evolution of the structure of a magnetic field with a vertical current sheet used in the standard flare model (Chen et al., 2015). This analysis showed that the following chain of physical processes can be realized during solar eruptions (e.g., CME), as outer perturbations are imposed on the system: 1) Plasma and magnetic fields on both sides of the current sheet move toward one another gradually, causing the current sheet to become thinner and thinner. 2) Magnetic reconnection commences gradually during this period producing two plasma jets inside the current sheet: one directed upward away from the solar surface and the other, downward toward the arcade of newly-reconnected magnetic loops. 3) The current sheet continually narrows until it becomes thin enough for the tearing mode and other instabilities to develop inside it. As a result, small-scale magnetic structures appear in the current sheet, inducing multiple *X*-shaped magnetic neutral points between each pair of the magnetic islands where fast magnetic reconnection can occur. The initially quasi-static current sheet then evolves into a non-uniform and dynamic state, when the reconnection outflows become highly intermittent. 4) The arcade of newly-reconnected, dense magnetic loops acts as an obstacle, which stops the high-speed reconnection outflows. The speed of the reconnection outflows gradually increases as the current sheet continually narrows down, and as soon as the outflows become super-magnetosonic, a standing fast-mode shock forms exhibiting itself as a TS.

The relationship between the radio sources and TS was demonstrated most convincingly in the work by Chen et al. (2014) based on the Very Large Array (VLA) data. The VLA radio images in the frequency range 1-1.8 GHz show the radio sources of the fast radio bursts coincide spatially with the sources of hard X-rays on the top of the loop. These observations show also that the disappearance of the radio bursts (spikes) coincides with the destruction of TS. Based on these data, it was concluded that the main source of accelerated electrons in the events discussed above are TS (as a result of drift acceleration in shocks).

According to the papers mentioned above, the main sign of the presence of TS are the peculiarities of the dynamic spectra of radio bursts, in particular, the absence of a frequency drift in the type II-like bursts, numerous short - lived and narrow frequency band bursts (spikes), and the appearance of new hard X-ray sources coinciding spatially with the radio sources. The scheme of TS formation agrees with the standard flare model including the reconnection of the magnetic field lines (see Fig. 1 in (Mann et al. 2009)). In this paper, we have used the data on two solar events of September 12, 2004 and November 18, 2003 to analyze the unusual radio bursts with small or absent frequency drift that coincide in time with hard X-rays sources. Section 2 describes the observations; the data are discussed and the mechanism of radio emission is described in Section 3; Section 4 presents a conclusion.

## 2. Observations

### 2.1 Event of September 12, 2004

The event was observed following the M4.8 flare (Fig. 2, *GOES* 12 XRA 1-8A, https://www.swpc.noaa.gov/products/goes-x-ray-flux) in NOAA AR 10672 (N04E42).
The type II-like burst occurred in the time interval 00:29:00 - 00:33:25 UT. The fragments of dynamic radio spectrum of this event in the range of 2.6–3.8 GHz of the spectrograph at the Huairou station (National Astronomical Observatory of China, Chinese Solar Broadband Radio Spectrometers at Huairou (SBRS/Huairou, Fu et al. 1995; 2004) are shown in Fig. 1.

The radio phenomenon included three separate radio bursts, similar to type II bursts with small frequency drift. The first burst occurred in the time interval 00:29:00 - 00:30:00 UT. On the whole, the event is a complex wave-like structure consisting of the separate blobs (duration up to 10 s and a bandwidth of ~0.1f) with an inner fine structure (short-lasting bursts with a duration of ~0.1 s and zebra-like structure near 00:29:30 UT). In this burst the frequency drift stops and changes its sign at frequencies of ~2.6 GHz. After 2.5 min, one can see this effect is repeated, and after 5 min, a new radio emission band appears with the frequency drift stopped at higher frequencies (~2820 MHz). The drifting bands in the time period 00:32;57 – 00:33:10 UT can be considered as the known band splitting of the type II bursts. It should be noted that radio emission in all these bursts was strongly polarized. On the hole, by the polarization characteristics and the parameters of the dynamic spectrum, this event bears a similarity to the chains of type I bursts in noise storms in the meter wave range.

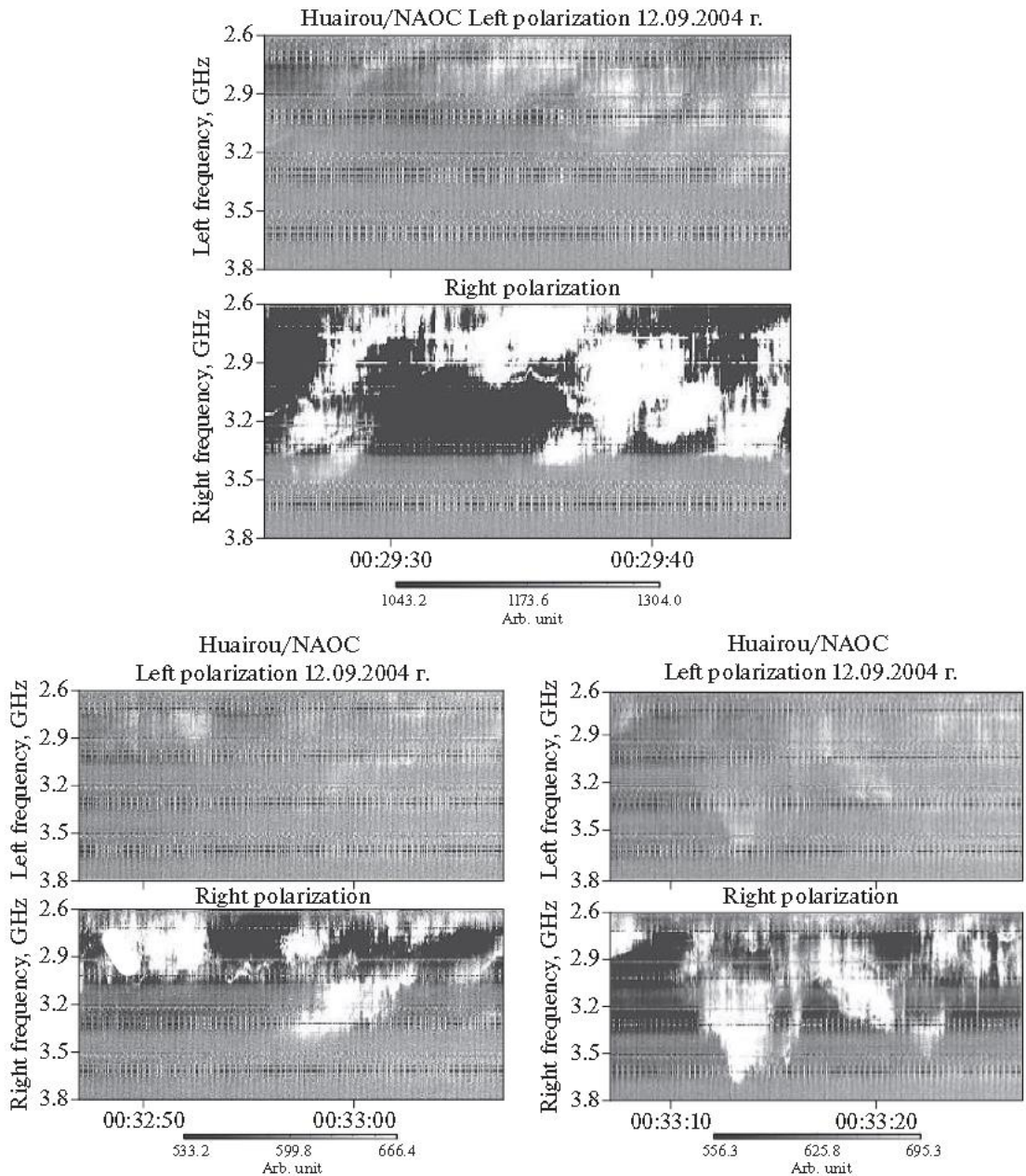

**Fig. 1.** Dynamic spectrum of the spectrograph of the National Astronomical Observatory of China, the SBRS/Huairou station in the left and right polarization in the microwave range of 2.6-3.8 GHz (Fu et al. 1995; 2004). The event of the September 12, 2004 with the radio burst showing change the frequency drift at frequencies of ~2.6 GHz.

From the first spectrum in Fig. 1, one can see the frequency drift stop and change sign at ~00:29:35 UT, near the maximum of the soft X-ray burst according to *GOES* data in Fig. 2. At that time, the sources of hard X-rays in the energy range of 25−50 keV (RHESSI data, Ramaty High Energy Solar Spectroscopic Imager (Lin et al. 2002; Hurford et al. 2002) were located at the foot-points of the loop above NOAA AR 10672, and the sources of emission in the energy range of 12−25 keV include also the source at the loop top (in comparison with the SOHO/MDI magnetogram). The RHESSI HXR time profiles in Fig. 2 show enhanced HXR emission in all the channels, from 3 to 100 keV, at the time of the radio burst. The maximum growth of HXR emission was observed in the channels 6–12 and 12–25 keV, and the channels 25–50 and 50–100 keV differed in the pulsed nature of the time profiles. Thus, in this event the radio feature appears at the start of the impulsive phase of new sources of HXR emission.

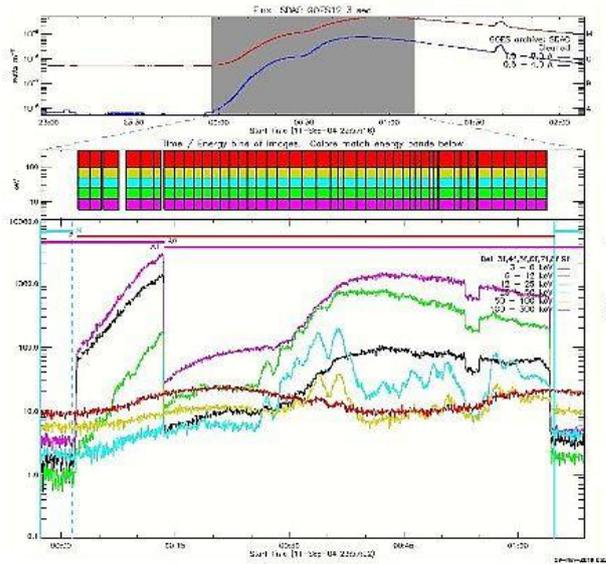
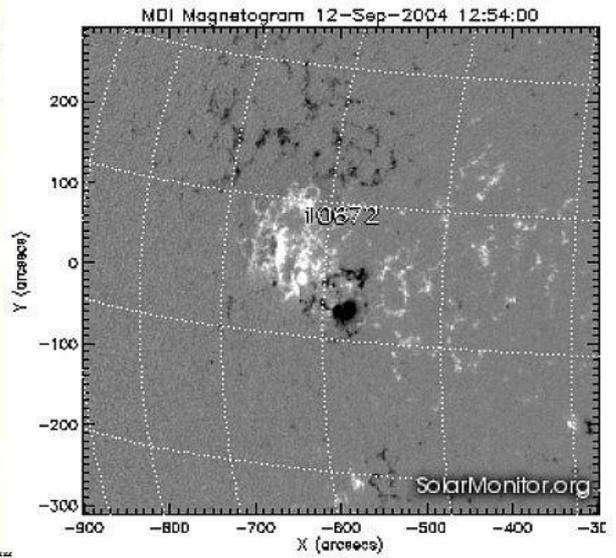
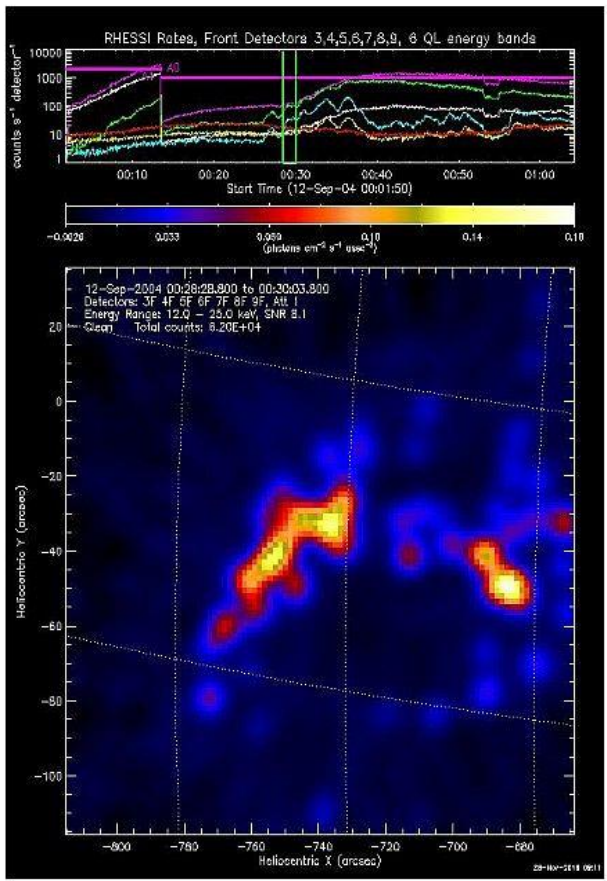
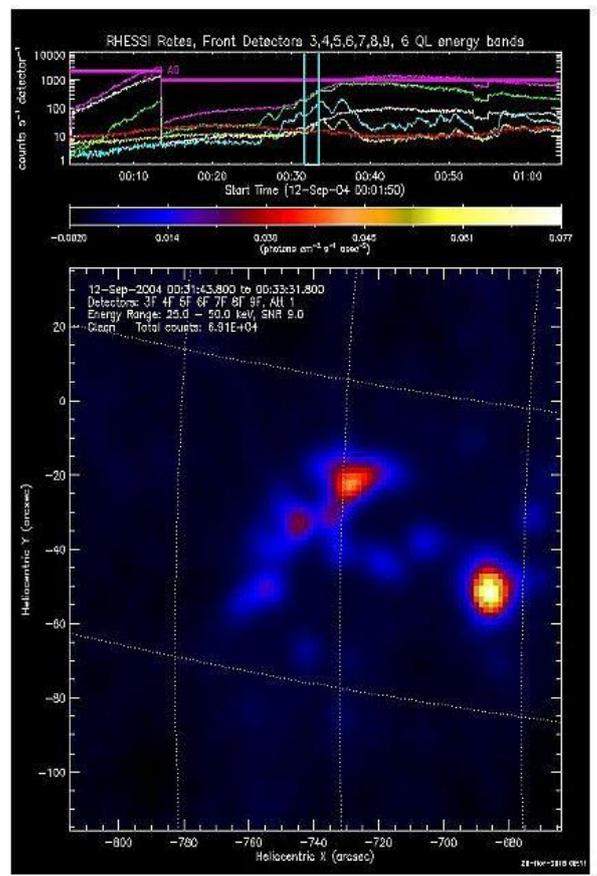

**Fig. 2.** SOHO/MDI magnetogram of NOAA AR 10672, GOES profiles and sources of HXR RHESSI of the event of September 12, 2004 at the beginning of the radio bursts 00:29-00:30 UT. Hard X-ray sources of 25-50 keV were located at loop bases above the main spots of AR, and in the lower-energy range of 12-25 keV we can also see a source at the loop top.
https://hesperia.gsfc.nasa.gov/rhessi_extras/flare_images/2004/09/12/20040912_0002_0104/CLEAN/movies/a_hsi_image_movie_20040912_0002_0103_12_25kev.html

*2.2. Event of November 18, 2003*

After a series of flares in the AR 10501 at the center of the solar disk, interesting radio bursts were observed following the M3.9 flare (08:12−08:31−08:59 UT).

The dynamic radio spectrum of the microwave burst recorded with the spectrograph of the SBRS/Huairou Station (Fig. 3a) and the Ondřejov Observatory (Jiřička et al. 2001) (Fig. 3b) similar to the dynamic spectrum of the type II burst at meter waves was observed in the time interval 08:24:50 - 08:25:30 UT. One can see a frequency drift of the emission bands to low frequencies, which stops at 2,5 GHz and changes its sign to positive. Again, it should be noted the blob structure of the emission bands.

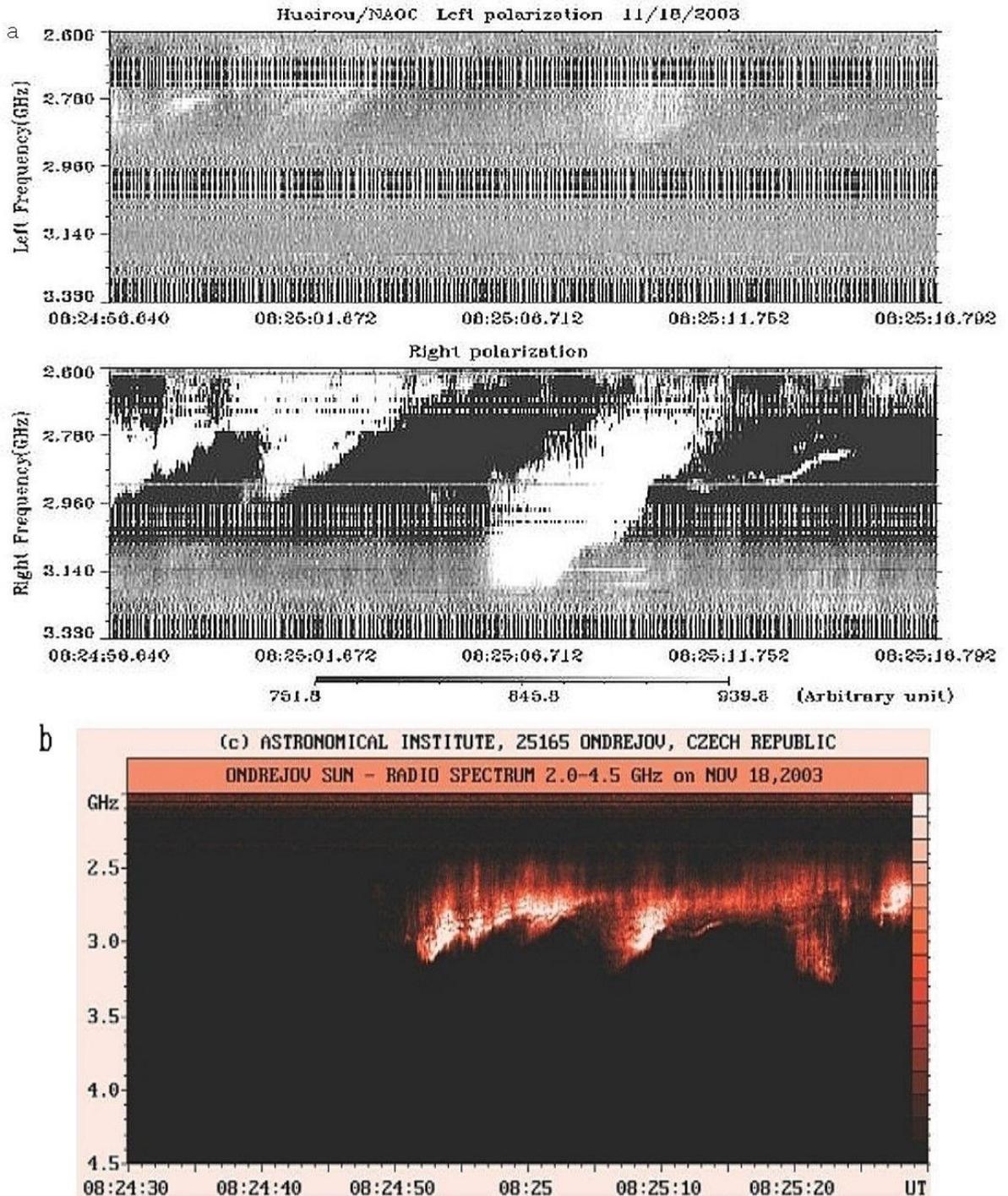

**Fig. 3. a)** Dynamic spectrum of the spectrograph at the Huairou station of the National Observatory of China in the left and right polarization in the microwave range of 2.6-3.8 GHz (Fu et al. 1995; 2004). b) Spectrum of the Ondřejov Observatory in the 2-4.5 GHz range (Jiřička, et al. 1993).

The interesting peculiarity of this event is that the blobs consisted of the drifting narrow-band filaments similar to zebra-structure in the type IV solar continual radio bursts. It should be noted that the burst occurred at the moment of maximum of the second GOES soft X- ray burst ( 08:25 UT) between the two CMEs that appeared on *SOHO*/LASCO C2 at 08:00 and 08:45 UT (two bottom panels in Fig. 4).

https://cdaw.gsfc.nasa.gov/movie/make_javamovie.php?date=20031118&img1=lasc2rdf and https://solarmonitor.org/index.php?date).

The RHESSI HXR time profiles in Fig. 4 show enhanced HXR emission in several channels, from 3 to 100 keV, at the time of the radio burst. The maximum growth of HXR emission was observed in the channels 6–12 and 12–25 keV, and the channels 25–50 and 50–100 keV differed in the pulsed nature of the time profiles (similar to the event on 12.09.2004).

On the RHESSI images (Fig. 4) we see the emergence of new sources: in the 12-25 keV channel - triple source, most probably, at the top of the loop, and in the 25-50 keV channel – source with coordinates (-30,-285 arcsec), probably at the base of the loop above the trail spot of AR (according to MDI magnetogram). The source with coordinates (-80,-325 arcsec) remained for a long time after the first maximum of the event around 07:50 UT.

Thus, during the radio burst on 18 November 2003 we see the emergence of new HXR sources in several energetic channels, and in this event the radio feature appears at the maximum phase of HXR emission in the energy channel 25-50 keV.

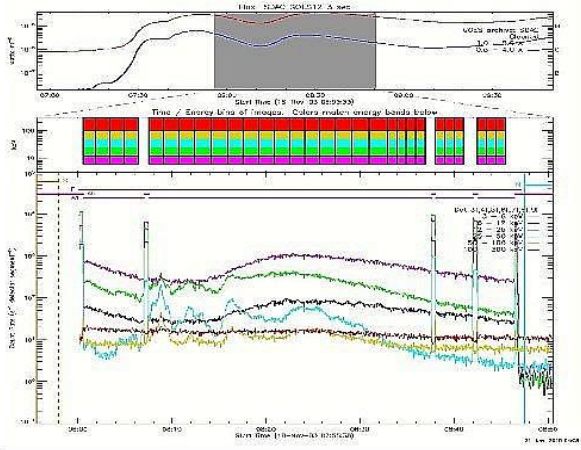
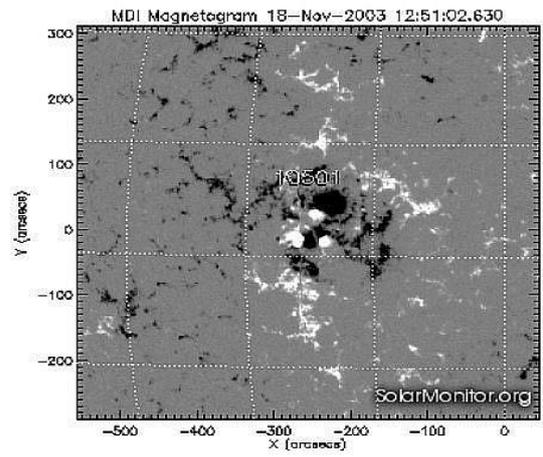
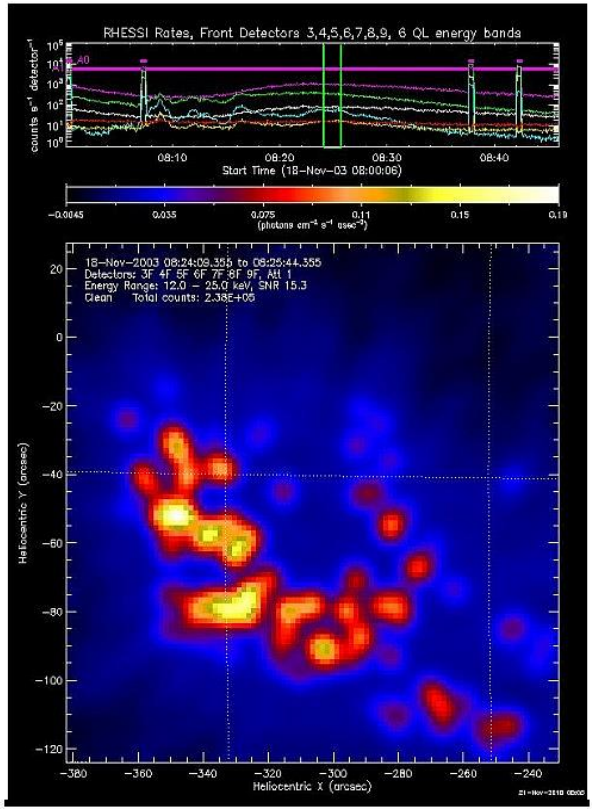
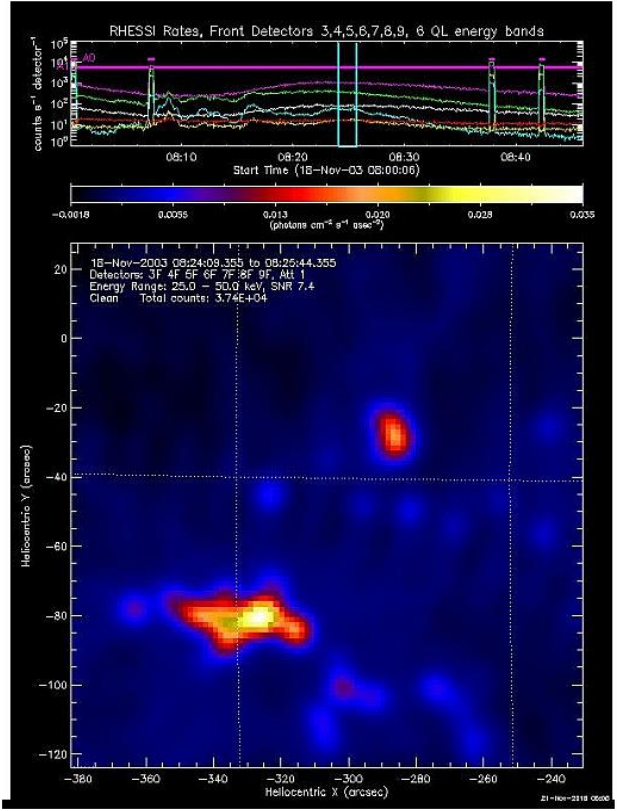
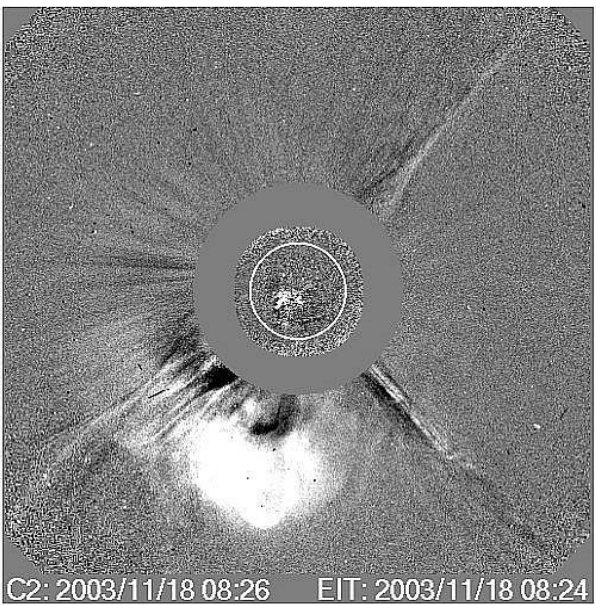
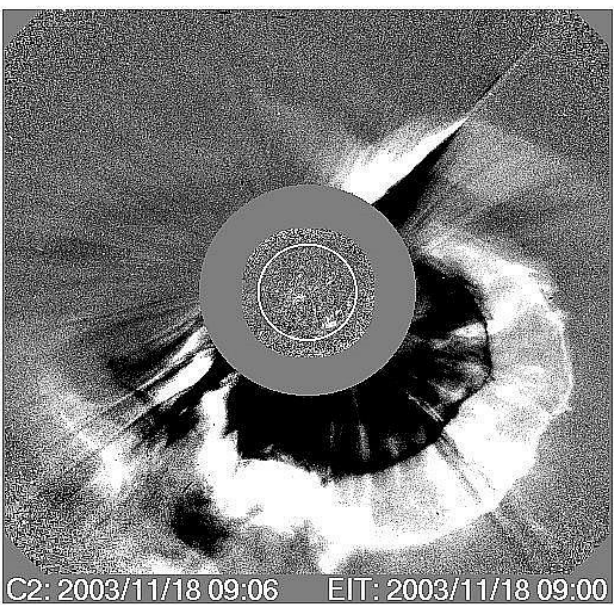

**Fig. 4.** SOHO/MDI magnetogram of NOAA AR 10501, GOES profiles, sources of HXR RHESSI and *SOHO*/LASCO C2 images of CMEs of the event of November 18, 2003. The HXR images are given at the time moment of the radio burst (Fig. 3).
https://hesperia.gsfc.nasa.gov/rhessi_extras/flare_images/2003/11/18/20031118_0800_0847/CLEAN/movies/a_hsi_image_movie_20031118_0800_0845_25_50kev.html

### 3. Discussion

The examples of a complex picture of electromagnetic emission in a broad wavelength range described above (X-rays, unusual radio bursts with the dynamic spectra similar to solar type II radio bursts or to chains of type I bursts, and spikes) urged various authors to suggest that accelerated particles might appear in the cusp-shaped magnetic structures as a result of various acceleration mechanisms and mechanisms of generation of electromagnetic emission, including the formation of termination (quasi- standing) shock waves (Mann et al. 2006, 2009; Warmuth et al. 2009; Aurass et al. 2013).

The currently discussed acceleration mechanisms may be divided into three groups: 1) Acceleration in large-scale electric fields arising, for example, during the reconnection of oppositely-directed magnetic field lines. This mechanism is widely used in many models of solar flares. 2) Stochastic acceleration where particles gain energy from turbulent waves. This mechanism is used to explain electromagnetic emission in a wide frequency range. 3) Shock acceleration. The shock waves are, usually, considered in connection with the coronal mass ejections (CME), and are used to explain the solar type II radio bursts and acceleration of particles in interplanetary space. All these acceleration mechanisms can be realized in the cusp-shaped (or helmet-shaped) magnetic structures in the solar atmosphere. As for the shock waves, they can be formed as "quasi-standing" or termination (TS) shock waves.

The TS parameters change because of variations in the parameters of the oncoming magnetic structures (islands) or outflows. Thereby, the shock drift acceleration, forming of the magnetic islands, and their interaction, i.e., the appearance of complex nonstationary MHD-structures in the helmet-like region in the solar atmosphere can result in generation of the electromagnetic emission in a wide spectrum, including radio bursts with unusual dynamic spectra.

*3.1 Event of September 12, 2004*

The observed behavior of X-ray sources (images and temporal history) and dynamic spectrum of the radio durst corresponds to the magnetic reconnection in accordance with the standard flare model at heights corresponding to plasma frequencies above 3300 MHz (<15000 km). The lower standing shock wave formed at approximately 0029 UT when the HXR emission began to enhance and the radio emission appeared. The lower TS formed due to the collision of the plasma flow emerging from the magnetic reconnection region with the magnetic loops above the loop top. The radio bursts had approximately the same frequency of ~2600 MHz where the frequency drift changed their sign from negative to positive. The termination shock wave remained an active source of accelerated electrons until 0035:45 UT, as judged by the fine structure (spikes and fibers) in the spectrum of the final burst shown in Fig. 1. By the time under discussion, it began to move downward (with the possibility of a break), because the drift of the final burst had stopped already at frequencies of ~2800 MHz.

One can also notice that in this event the radio burst show a concurrency with the HXR (RHESSI) count rates at the start of the impulsive phase of HXR emission.

*3.2. Event of November 18, 2003*

This event was accompanied by a powerful CME and the interesting radio-burst coinciding in time with the soft X-ray maxima (GOES X-rays). This phenomenon involves the drift of the emission bands to low frequencies, a stop of the drift at a frequency of 2.5 GHz, and the subsequent reversal to the positive drift. The change of frequency drift of the emission bands coincides with the soft X-ray maximum at 08:25 UT (Fig. 4). The source of hard X-rays at 08:10 UT was located at the top of the flare loop. Later on, at 08:14 UT, the sources of the higher-energy 12-25 keV emission (RHESSI) are already located at the loop foot points, if compared with the MDI magnetogram. The emergence of new HXR sources implies new acts of electron acceleration. Fig. **4** shows also a concurrency of the hard X- ray burst with the radio burst. Fast electrons accelerated in the TS and propagating downward leaded to the enhancement of hard X-ray at the foot points of the loop. But unlike the previous event of September 12, 2004, in this event the type II – like radio burst shows a concurrency with the HXR (RHESSI) count rates at the maxima phase of HXR emission. Both the behavior of X-ray sources and the appearance the type II – like radio burst are consistent with the standard model of a flare with magnetic reconnection and the forming lower TS above the magnetic loop system.

*3.3. Mechanism of Radio Emission*

In all the phenomena, the radio bursts similar to metric type II bursts display a stop and reversal of the frequency drift. Analysis of all available data on the corresponding flares indicates the formation of standing shock waves (TSs), in the fronts of which, a drift acceleration of electrons occurred as confirmed by generation of new HXR sources and fast radio bursts (spikes), fibers, and zebra structures. The sources of the emission bands with a small or absent frequency drift were most probably connected with the lower TSs and located above the flare loops. All phenomena are characterized by the presence of various fine-structure elements (mainly, spikes, fibers, and zebra-structures) generation of which is due to appearance of the electrons accelerated in the front of a standing shock wave.

The standing shock waves described above have so far been regarded the fast electron accelerators (drift acceleration).

According to Mann et al. (2006), such the accelerated electrons could be a source generating one of the peculiar radio burst on October 28, 2003 with fibers in the decimeter range and a stop of the frequency drift recorded with a Chinese spectrograph.

However, the standing shock waves can also contribute significantly to the generation of radio bursts (namely, emission bands in the radio range that look like type II bursts), their internal structure being taken into account. Indeed, if the number of collisions in plasma is sufficiently small (as occurs in the solar atmosphere - chromosphere and corona), the shock front is an oscillatory structure consisting of a sequence of solitons of compression or rarefaction (Sagdeev, 1964). The type and scale of this structure depend on the Alfvenic Mach number $M = V_{sw}/V_A$ (here, $V_{sw}$ is the shock wave speed and $V_A = H_0/(4\pi N_0 m_i)^{1/2}$ is the Alfven velocity), the angle $\theta$ between the front plane and the direction of undisturbed magnetic field $H_0$, and the parameter $\alpha = H_0^2/4\pi N_0 m_e c^2 = \omega_{H0}^2/\omega_{L0}^2$, where $\omega_{H0} = eH_0/mc$, $\omega_{L0} = (4\pi e^2 N_0/m_i)^{1/2}$, $N_0$ and $H_0$ – electron density and strength of the upstream magnetic field, correspondently, and $c$ is the speed of light. Here the value $\theta = 0$ corresponds to transverse propagation of the shock wave relatively the upstream magnetic field, and $\theta = \pi/2$ corresponds to longitudinal propagation. As shown by Zaitsev (1970), shock waves differ qualitatively at $\theta < \theta$ cr and $\theta > \theta$ cr (sin $\theta$ cr = $V_{sw}/c$). The shock transition occurs through a sequence of solitary compression waves at $\theta < \theta$ cr and through a sequence of rarefaction waves at $\theta > \theta$ r. The characteristic scale of a solitary wave, e.g., for compression waves at $\theta \approx 0$ (quasi-transverse propagation) is $\delta \approx c/\omega_{L0} (1 + \alpha + \alpha^2 m_e/m_i)/(1 + \alpha m_e/m_i)^{3/2}$. The inhomogeneity of the magnetic field in solitons makes the electrons drift in the plane of the shock front. The relative drift velocity of electrons and ions $V_*$ = $4 \cdot 2^{1/2} \cdot 3^{-3/2} c \omega_{H0} (\omega^2_{L0} + \omega^2_{H0})^{-1/2} (M - 1)^{3/2}$ at M > Mcr = $1 + [8\pi N_0 kT(1 + \alpha) H_0^{-2}]^{1/3}$ (where k = $1.38 \times 10^{-16}$ erg/K is the Boltzmann constant) (Zaitsev, 1969) exceeds the thermal electron velocity $V_{Te} = (kT/m_e)^{1/2}$. This leads to the generation of plasma oscillations (the Buneman instability) and, as a result, to the generation of electromagnetic radiation. This mechanism made it possible to develop a comprehensive model of the source of the classical type II bursts (Zaitsev, 1977) and to explain the fine structure of radio bursts of these types generated in some phenomena in the form of harmonics of the electron plasma frequency with high numbers (*n* > 2) (Fomichev et al., 2013, 2018). According to (Zaitsev, 1977), the maximum power of radio emission from the unit surface of a source (e.g. for the second harmonic) is P ≤ $2 \cdot 10^{-7} (V/V_{Te})^3 N_0$ *erg/cm²s*. In the solar corona $N_0 \approx (10^7 - 10^8)$ cm$^{-3}$ and $(V/V_{Te}) \approx 3$-5, we have P ≤ (4- 400) *erg/cm²s*, which is sufficient to explain the radio-burst intensity observed. This conclusion also hold for the higher densities $N_0 \approx (10^9 - 10^{10})$ cm$^{-3}$ in active regions and reconnection outflow jets expected and actually derived from EUV, HXR and radio observations (P ≤ (400- 40000) *erg/cm²s*).

One can see from the previous expression that the critical Alfvenic Mach number decreases with increasing magnetic field and reaches the value Mcr = $1 + \frac{3}{4} (2 V_{Te}^2/c^2)^{1/3}$ at α >> 1. Under these conditions, quasi-transverse shock waves as possible sources of type-I burst chains were analyzed by Zaitsev and Fomichev (1972).

The laminar structure of shock waves moving along the magnetic field in a plasma with β = $8 \pi N_0 kT / H_0^2$ << 1 exists in the range 1 < M < 1.5. It has the form of a spinning spiral with a step δ ≈ $c/\omega_{Li}$ and the magnitude of the transverse field behind the front $H_\perp = 2^{1/2} H_0 (M^2 - 1)^{1/2}$ (Kurtmullaev et al., 1971). The front of the shock wave also involves a drift current directed across the disturbed magnetic field. However, the relative velocity of electrons and ions $V_d$ in that case is much lower than the electron thermal velocity $V_{Te}$; therefore, the ordinary Buneman instability occurring in transverse shock waves does not develop. Instead, a modified Buneman instability with a very low excitation threshold $V_d > V_{Ti}$ develops in longitudinal shock wave at $Te \approx Ti$. The excitation condition begins to be satisfied in front sections with $H_\perp = 2 H_0 (M - 1)$ at Alfvenic Mach number M > $1 + (V_{Ti}/2 V_A)^{1/2}$ (Zaitsev and Ledenev, 1976). As a result, longitudinal waves are excited at the lower hybrid frequency ω ≈ $(\omega_{He} \omega_{Hi})^{1/2}$ with an energy density of about $W \approx N_0 m_e V_d^2$ in the saturation mode (Galeev and Sagdeev, 1973). The energy of these waves is transferred to particles; i.e., the plasma at the shock front is heated. It follows from discontinuity relations for a longitudinal shock wave that the kinetic pressure behind the shock wave front can reach the pressure of an undisturbed magnetic field (Zaitsev, 1977). This means that electrons are heated to the temperature $Te \approx Ti \approx H_0^2/ 8\pi N_0$, i.e., acquire an average thermal velocity $V_t \approx \frac{1}{2} (m_i/m_e)^{1/2} V_A$. In the solar corona, $V_A \approx 5\cdot10^7 - 10^8$ cm/s; therefore, $V_t \approx$ (1 - 2) $10^9$ cm/s significantly exceeds the thermal velocity of electrons in the undisturbed corona ($V_{Te} \approx 5\cdot10^8$ cm/s), and the concentration of fast electrons escaping beyond the front can reach $Ns \sim \frac{1}{2} (m_e/m_i)^{1/2} N_0 \approx 10^{-2} N_0$ according to the conclusions made by Zaitsev and Ledenev (1976). In other words, the front of a longitudinal shock wave at β << 1 and at sufficiently large Mach numbers is an emitter of fast electrons, which, falling into a cold plasma behind the front, can excite coherent plasma waves and be a source of radio emission.

Under real conditions, when the standing shock wave forms as a result of incidence of a fast plasma stream on a loop arcade, it may have the features of either transverse or longitude· shock in different parts of the front. Excitation of turbulence in shock waves requires that the Mach number exceed a certain critical value Mcr (see expressions above): at the adopted parameters of the solar corona, Mcr ≈ 1.2 for a longitudinal shock wave, and Mcr ≈ 1.05 in the case of α >>1 ($\omega_{H0} >> \omega_{L0}$) and Mcr ≈ 1.16 in the case of α << 1 ($\omega_{H0} << \omega_{L0}$) for a transverse shock wave. With the parameters for the decimeter range $N_0 = 10^{10}$ cm$^{-3}$ (plasma frequency $fp$ = 900 MHz), temperature $Te = 5\cdot10^6$-$10^7$K, and $H_0 = 10$ G, we obtain Mcr = 1.36-1.7. For the microwave range $N_0 = 10^{11}$ cm$^{-3}$ ($fp$ = 2846 MHz) and $H_0 = 50$ G, we obtain Mcr = 1.52-1.9. It should be noted that such high temperatures are usually relevant to the X-ray sources, temperature of the undisturbed solar corona is about (1-2)$10^6$ K, and values Mcr < 1.4. Such values can be easily exceeded. In addition, since the incident flow parameters are variable, the standing shock waves or their individual segments will be variable and flickering radiation sources with a characteristic time that depends on the speed of the shock wave and the size of inhomogeneities in the corona. For example, a shock-wave discontinuity may be caused by fluctuations in the magnetic field direction with respect to the shock wave speed and other factors. A slight deviation of Θ from transverse makes the conditions for the excitation of plasma waves in the shock front more stringent, and at $\Theta \geq V$sw$/c \approx 0.033 \approx 2°$, the generation of plasma waves stops almost completely. It means that in the real coronal conditions there are many physical reasons (e.g. the dynamic plasma structure in the reconnection out flows, limitations on the generation of plasma waves into the shock front, the magnetic islands impinge on shock front and a temporary disruption of the termination shock surface) for the dynamic spectrum of the type –II like radio burst to have an interruptive and a blob structure.

Moreover, it is necessary to note that in the model discussed here the physical drivers for generation of HXR emission and type II – like radio bursts are different.  Nonthermal HXR emission is generated by electron beams typically resulted from shock drift acceleration at shocks, and HXR sources are located in the magnetic loops or at the loop top, that is out of  the shock front. The type II - like radio burst in scenario here is due to the generation of plasma oscillations (the Buneman instability) and, as a result, to the generation of electromagnetic radiation within a shock front. In may be concluded that the electron populations for HXR and radio emission in this case are not the same. Therefore, the appearance of radio features may occur at different phase of HXR emission. Such a situation we observe in our case, the radio feature in the event of September 12, 2004 appears at the start of the impulsive phase of HXR emission, and in the event of November 18, 2003 the radio feature appears at the maximum phase of HXR emission.

As to the fine structure of the blobs of radio emission bands similar to zebra-structure in the type IV solar continual radio bursts, a number of mechanisms were suggested. The most widely discussed models are the model when generation of plasma waves occur in the time of coincidence of the frequency of upper hybrid resonance with the frequency of cyclotron harmonics (Double Plasma Resonance, Zheleznyakov et al. 2016), and the model of zebra patterns during the interaction of plasma waves with whistlers (Chernov, 2011). In any case, for generation of zebra-structure an anisotropic distribution of fast electrons particles is needed. Possibility of realization of this condition in TS requires a special analysis.

## 4. Conclusion

We have considered two events with radio bursts in the decimeter and microwave ranges that are similar to the type II meter bursts. The radio bursts analyzed are characterized by a stop of the frequency drift and even its reversal. The analysis of all the data available on the corresponding flares indicates that the quasi-standing shock waves could be formed (with the drift acceleration of particles taking place at their fronts). The estimates of the possible Alfvenic Mach numbers of the flare-generated shock waves show that they can quite probably be formed in the flare plasma of the solar corona. The events with a strongly polarized radiation (such as the event of September 12, 2004) are associated with the lower standing shock wave (i.e., near the top of the loop structure), where the magnetic field can be sufficiently large ($\omega_H \geq \omega_L$).

In all the phenomena considered above, the dynamic spectrum of the type II –like radio bursts have an interruptive and a blob structure. This can be explained by the fact that the characteristics of the shock wave changes during propagation of the shock wave in the inhomogeneous solar atmosphere, e.g. due to fluctuations in the magnetic field direction with respect to the shock wave front from transverse to quasi-longitudinal. For transverse shock waves the generation of radio emission is due to development of the Buneman plasma instability in the shock front. For longitude shock waves, a more efficient generation mechanism is associated with a high turbulence of low-frequency waves behind the shock front. For further analysis of the observed unusual radio bursts, one should also take into account the contribution of possible packets of MHD waves (such as Alfven and fast magnetosonic waves) as sources of additional electron acceleration, their dynamics, and radio emission properties (quasi-periodic oscillations).

It should be noted that the generation of phenomena in the post-eruptive phase of solar flares indicates that they are associated with processes in the helmet-shaped magnetic structures, which can combine a large number of physical processes, such as magnetic reconnection, particle acceleration, the formation of magnetic islands, their interaction each with other and with underlying arch structures, and the formation of standing shock waves.

**Acknowledgments**

The authors are grateful to the referee for constructive comments and help in updating RHESSI observational data coinciding in time with the discussed radio bursts.

The authors are grateful to the Chinese colleagues who kindly provided theirs spectral data obtained at the Huairou National Observatory (China) and to Dr. M. Karlicky for the spectrum of the Ondřejov Observatory. The authors are also grateful to the RHESSI, GOES, and LASCO teams for providing us with open access to relevant data. We would like to thank the referee for the valuable comments and suggestions. This work was supported by the Ministry of Science and Higher Education, project no. KP 19-270.